# Determination of Oxygen Saturation and Photoplethysmogram from Near Infrared Scattering Images

Ri Yong-U,    Pyon Young-Hui,    Sin Kye-Ryong [*]

(*Faculty of Chemistry, **Kim Il Sung** University, Pyongyang, DPR of Korea*)

* E-mail: ryongnam9@yahoo.com

**Abstract:** The near infrared scattering images of human muscle include some information on bloodstream and hemoglobin concentration according to skin depth and time. This paper addressed a method of determining oxygen saturation and photoplethysmogram from the near infrared (NIR) scattering images of muscle. Depending on the modified Beer-Lambert Law and the diffuse scattering model of muscular tissue, we determined an extinction coefficient matrix of hemoglobin from the near infrared scattering images and analyzed distribution of oxygen saturation of muscle with a depth from the extinction coefficient matrix. And we determined a dynamic attenuation variation curve with respect to fragmentary image frames sensitive to bloodstream from scattering image frames of muscle with time and then obtained the photoplethysmogram and heart rate by Fourier transformation and inverse transformation.

This method based on the NIR scattering images can be applied in measurement of an average oxygen saturation and photoplethysmogram even in local region of optically heterogeneous muscle and skin.

**Introduction**

Technologies measuring oxygen in human muscle are widely used in correct evaluation of physiological ability of athletes to organize scientifically training.[3] In general, measuring oxygen in muscle is based on measurement of NIR absorption and scattering in tissue.

The common muscle oximeters have single sensor or multiple sensors with a light source or several light sources or several detectors and the distance between source and detector (separation) can be controlled depending on a measured object.[4,5] Therefore, such sensors can't measure an attenuation change with separation at once and can't obtain the two-dimensional information of bloodstream in muscular tissue. In fact, an average measurement on a local region of muscle has significance because the surface of skin and muscular tissue are optically heterogeneous.

In our previous paper, it was theoretically proved that by using only one NIR scattering image of muscle it can be analyzed the attenuation change with separation distance because every pixel from a NIR CCD camera can work as a detector.[2]

Therefore, it seems that the NIR scattering image of muscle contains the same information as the repeated measurements on attenuation with various separation distances. But there is no report on the reliable method on the quantitative interpretation of this interesting NIR scattering image up to now.

In this paper, presented was a novel method based on the modified Beer-Lambert Law and diffuse scattering model for determining oxygen saturation and photoplethysmogram from the NIR scattering images of human muscle.

# 1. Determination of extinction coefficient matrix and analysis of attenuation change with separation distance from NIR scattering image of muscle.

## 1.1 Calculation and analysis of extinction matrix.

In scattering medium, Beer-Lambert Law is as follows:[6]

$$A(\lambda) = \mu_a(\lambda) \cdot L(\lambda) + G(\lambda) \qquad (1)$$

where $A(\lambda)$ is an attenuation of light with a given wavelength and expressed as logarithm of reciprocal of pixel intensity in case of the scattering image. $\mu_a$ is a extinction coefficient of medium, $G(\lambda)$ is a light attenuation by purely scattering in tissue, $L$ is an average path length of light and expressed as $L = DPF \cdot \rho$. $DPF$ (differential path factor) depends on wavelength, kind of tissue and geographical configuration of sensor.

In an attenuation image of muscle, attenuation of i-th pixel $a_i(\lambda)$ is given as follows:

$$a_i(\lambda) = \mu_{ai}(\lambda) \cdot l_i(\lambda) + g_i(\lambda) \qquad (2)$$

where $\mu_{ai}$ is an extinction coefficient of muscular tissue, $g_i$ is light attenuation by pure scattering in tissue, $l_i$ is an average path length of light and expressed as $l_i = DPF \cdot r_i$ ( $r_i$ is separation distance).

Therefore Eq. 2 can be written as follows:

$$a_i / r_i - g_i / r_i = \mu_{ai} \cdot DPF_i \qquad (3)$$

To obtain an extinction coefficient matrix on the measured image by using Eq. 3, it is necessary to know $R$, that is, separation distance matrix (SDM), corresponding to attenuation matrix. SDM can be obtained by replacing each pixel value of attenuation image with separation distance value measured with an incident point of light source as original point. Element of SDM corresponding to pixel coordinates m, n can be determined by Eq. 4.

$$R(m,n) = d \times [(m - x_0)^2 + (n - y_0)^2]^{1/2} \qquad (4)$$

where d is an physical distance (cm) between adjacent pixels, $x_0, y_0$ are pixel coordinates of incident point of the light source. When there is an incident point of light source out of the image to be analyzed, coordinates of incident point can be calculated by extending pixel coordinates of the image on the same scale.

Assuming that DPF and scattering factor are constant on a given muscular tissue, Eq. 3 can be written as follows:

$$a_i / r_i - g / r_i = a_i' / r_i = \mu_{ai} \cdot DPF \qquad (5)$$

where $a_i'$ is modified attenuation by correcting a base line. Denoting relative extinction coefficient in each pixel by $m_i = \mu_{ai} \cdot DPF$, extinction matrix M can be calculated by dividing attenuation matrix A where the base line was corrected by SDM.

Matrix M indicates distribution of extinction coefficient in the scattering image. If extinction coefficient matrix with respect to two wavelength($\lambda_1, \lambda_2$) was obtained, by using Eq. 6 and Eq. 7 the distribution of oxygen saturation can be calculated.

$$m_i = (\varepsilon_{HHb} C_{HHbi} + \varepsilon_{O_2Hb} C_{O_2Hbi}) \cdot DPF \qquad (6)$$

$$sp_i = \frac{-\varepsilon_{HHb}(\lambda_1) + rm_i \cdot \varepsilon_{HHb}(\lambda_2)}{\varepsilon_{O_2Hb}(\lambda_1) - \varepsilon_{HHb}(\lambda_1) - rm_i(\varepsilon_{O_2Hb}(\lambda_2) - \varepsilon_{HHb}(\lambda_2))} \qquad (7)$$

where $rm_i = m_i(\lambda_1) / m_i(\lambda_2)$ and $sp_i$ is i-th element in oxygen saturation matrix. And $\varepsilon_{HHb}(\lambda), \varepsilon_{O_2Hb}(\lambda)$ are molar absorption coefficients of $HHb$ (deoxyhemoglobin) and $O_2Hb$ (oxyhemoglobin), and $C_{HHbi}, C_{O_2Hbi}$ are concentrations of $HHb, O_2Hb$ in i-th pixel.

To obtain NIR images, here we used a home-made scattering image sensor with NIR light source and CCD camera in Fig. 1(a). By using this sensor, scattering images from human arm muscle were measured on two wave lengths (740nm, 850nm) respectively and transformed into gray intensity image $I$ to calculate an attenuation image A according to $A = -\log_{10}(I)$.

From the scattering image measured on comparatively homogenous muscle, we obtained the contour of attenuation and determined pixel coordinates of optical center (incident point) of two light sources from that contour, and then calculated SDM by Eq. 4. Centers of two light source are not the same, so their SDM are not the same. The images shown in Fig. 1(b) and Fig. 1(c) are those that

SDM was normalized with its maximum and minimum. In the measurements of the images, the physical separation distance between adjacent pixels is 0.0032cm and the size of measured region is 1.53cm×2.05cm.

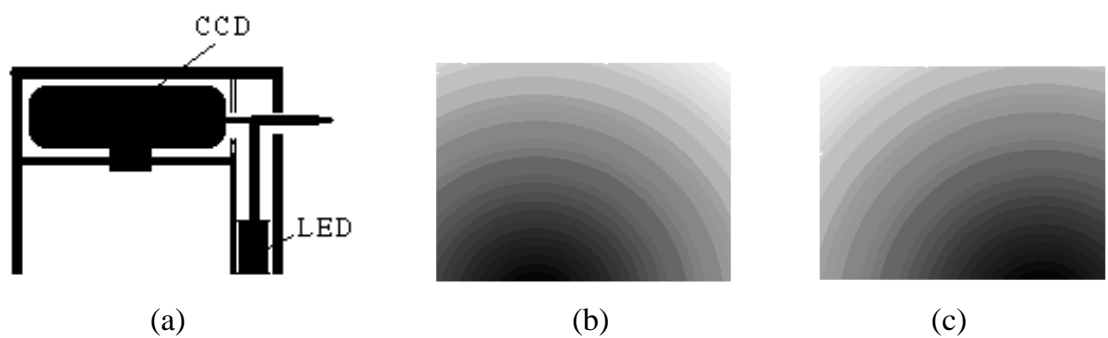

Fig. 1 Sensor's view and separation distance distribution for two light sources:
(a) scattering image sensor, (b) 740nm light source, (c) 850nm light source

The extinction coefficient image calculated by Eq. 5 from the acquired attenuation image and separation matrix was shown in Fig. 2.

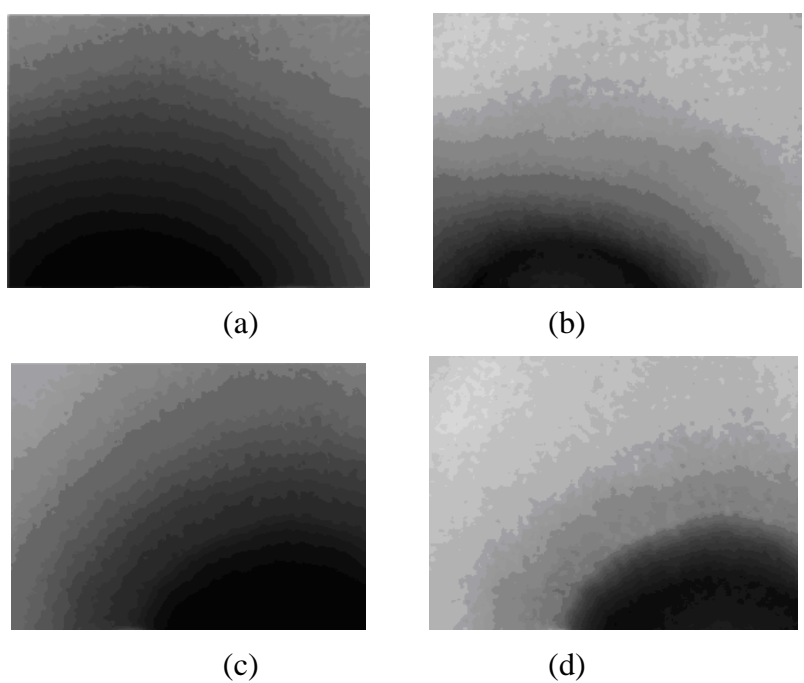

Fig. 2 The acquired attenuation image and extinction coefficient image.
(a) attenuation image acquired at 740nm, (b) extinction coefficient image calculated from (a),
(c) attenuation image acquired at 850nm, (d) extinction coefficient image calculated from (c).

In the extinction coefficient image the actual extinction coefficient value is as small as 1/10 of pixel value.

From the relative extinction coefficient matrix with 2 wavelengths, oxygen saturation matrix were calculated by using Eq. 7. The distribution of oxygen saturation values from oxygen saturation image can be shown in a histogram in Fig. 3.

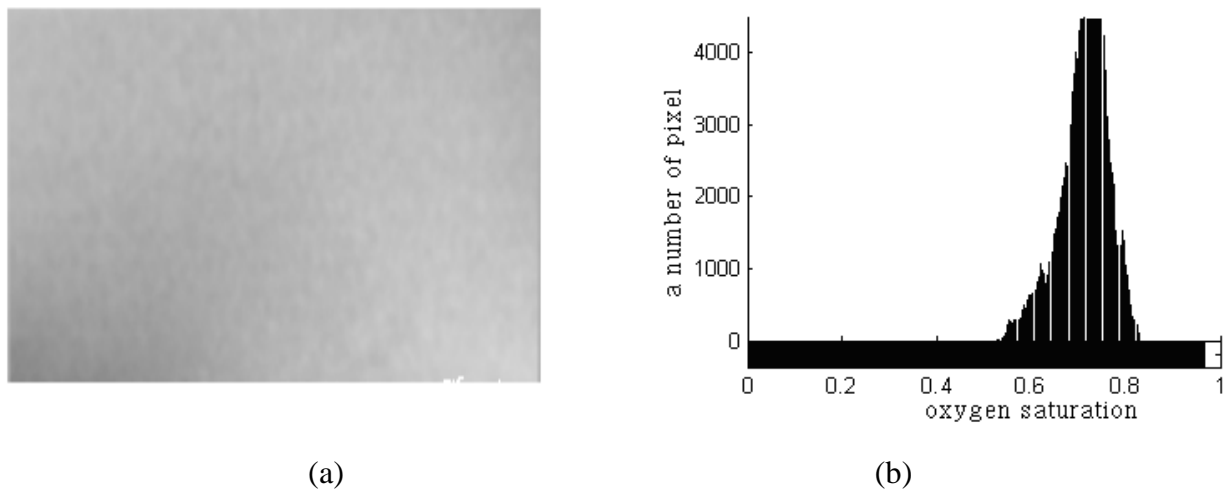

(a)          (b)

Fig. 3    Image on oxygen saturation (a) and distribution of oxygen saturation value (b)

The oxygen saturation has a peculiar distribution according to the measured local region and the optical anisotropy of muscular tissue can be evaluated from it.

**1. 2    Changes of attenuation and extinction coefficients according to separation distance.**

Extinction coefficient image from scattering image doesn't indicate the distribution of extinction coefficients in muscle surface itself, but it indicates changes of extinction coefficient with various separation disatnce. That is, change of extinction coefficients in muscle layer is shown according to depth from skin. According to the semi – infinite medium model, thickness of muscle layer determined from the statistical path of banana – shaped scattering light is about half of separation distance.[7]

By attenuation matrix (or extinction coefficient matrix M), change of attenuation (or extinction coefficient) with separation distance (depth of muscle) can be analyzed. Because there are many pixels with the same separation distance in attenuation matrix A, the average attenuation value of the pixels can be obtained by averaging the attenuation value.

By interpreting the average of the attenuation value and its standard deviation, it can be desired to evaluate an optical heterogeneity of the measured muscular tissue.

To obtain distribution of attenuation values and extinction coefficient values with separation distance, the scattering pictures where every point is corresponding to a pixel from the attenuation image and extinction coefficient image were shown in Fig. 4.

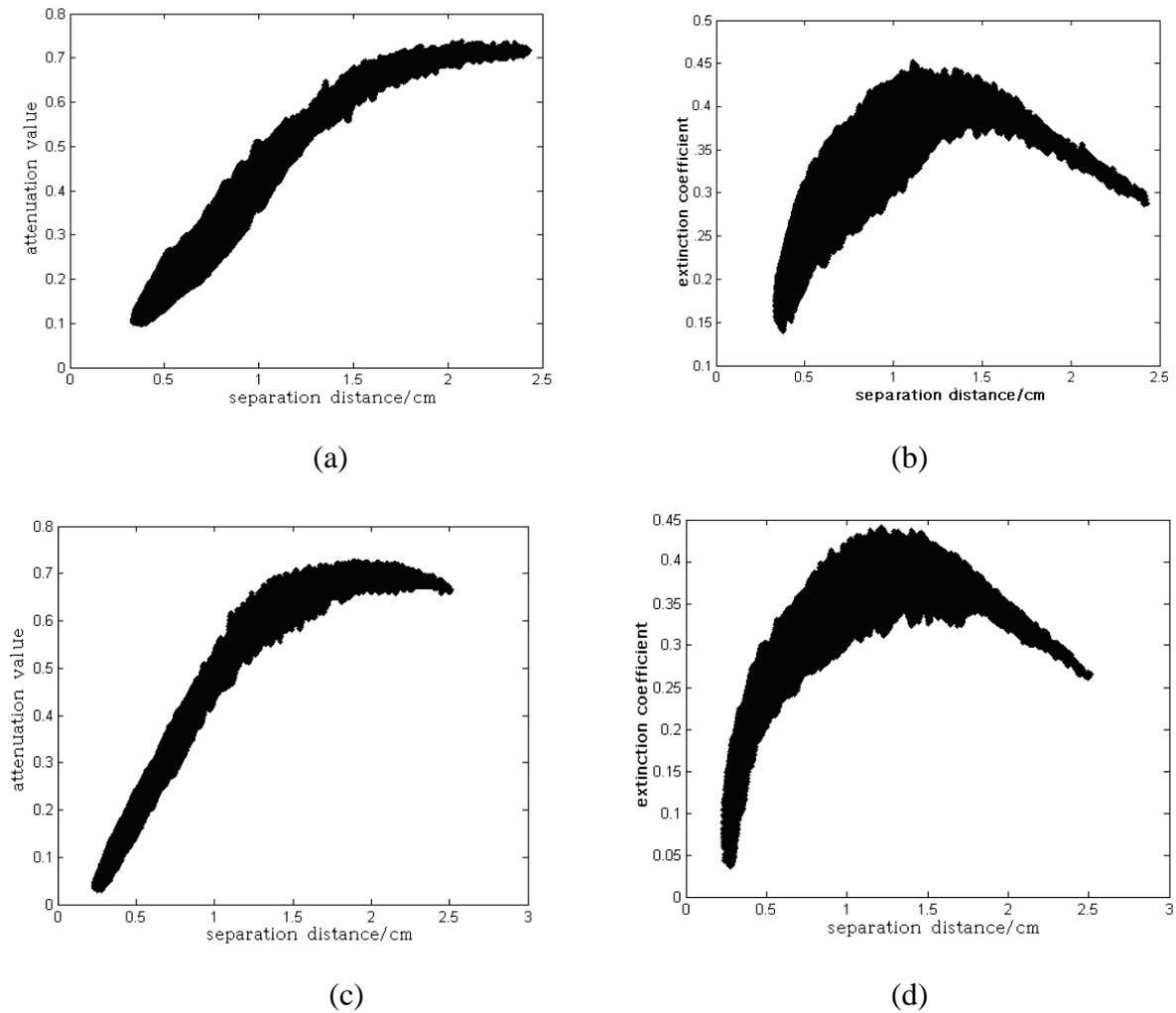

Fig. 4 The distribution attenuation and extinction coefficients with separation

(a) attenuation at 740nm, (b) attenuation at 850nm,

(c) extinction coefficient at 740nm, (d) extinction coefficient at 850nm

From Fig. 4, it can be seen that attenuation and extinction coefficient values depend on the separation distance and the direction of separation vector showing the direction from the origin point to pixels and the dependency is concerned with anisotropy of muscular tissue and a structure of image sensor. In case of using the same sensor the dependency is entirely related to the anisotropy of muscular tissue and analysis of change width of attenuation in a certain separation distance makes it possible to evaluate the anatomical structure of the measured muscle.

The average attenuation and extinction coefficients on each separation were calculated by averaging the attenuation values and extinction coefficients in the pixel coordinates of attenuation image and extinction coefficient image corresponding to pixel coordinates of each contour in the separation distance image. And the standard deviation of the extinction coefficient value in each separation was calculated and used as an index to evaluate anisotropy of the muscular tissue.

The changes in average attenuation and extinction coefficient according to separation was shown in Fig. 5 and the changes in standard deviation of extinction coefficient in Fig. 6.

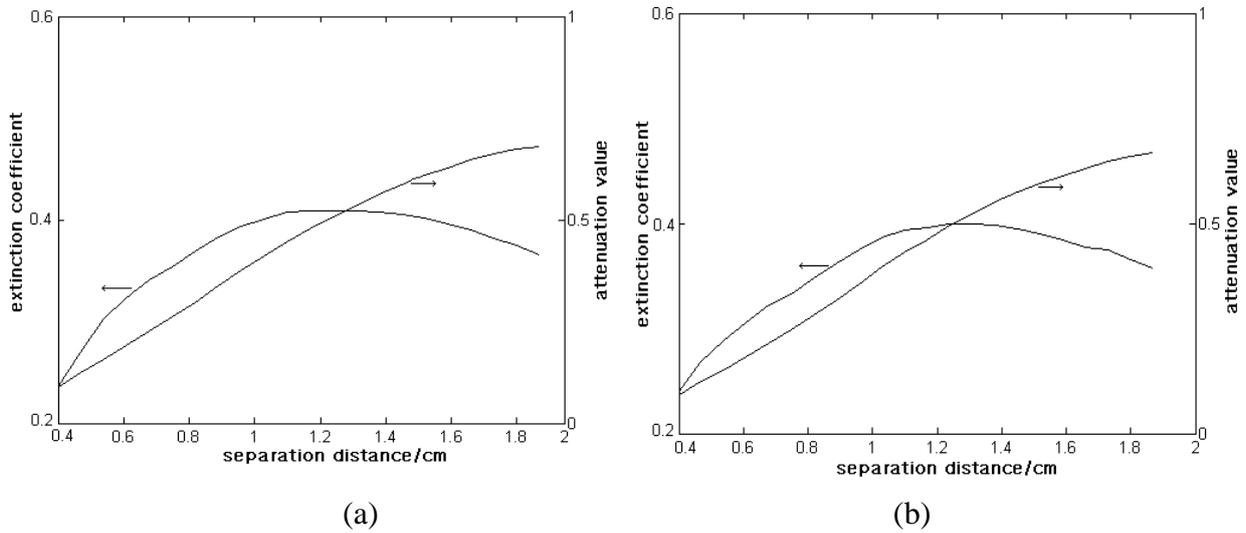

Fig. 5  Change in average attenuation and extinction coefficient according to the separation

(a) 740nm,    (b) 850nm

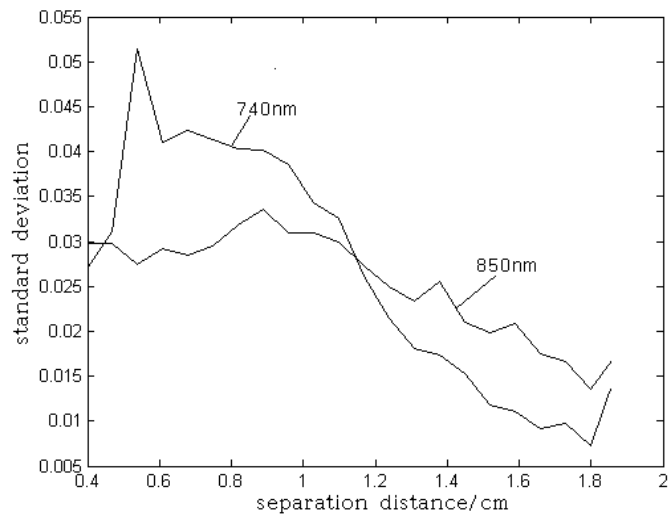

Fig. 6  Change in standard deviation of extinction coefficient according to the separation.

As shown in Fig. 6, standard deviation of the extinction coefficient is so large in the short separation because the heterogeneous surface of muscle was reflected more clearly in the short separation. And in the long separation, it can reflect the properties of muscle in depth.

Therefore, the standard deviation curve can be used as an index to reflect the muscular property.

## 2. Determination of variation curve of blood stream from NIR scattering image of muscle.

In the ordinary pulse oximeters for muscle, the photoplethysmogram was obtained by measuring the temporal change of penetrating light just in peripheral muscle such as tip of finger or earlobe. Therefore such an oximeter cannot measure the photoplethysmogram in various parts of muscle.[1,8-10]

During systole and diastole of heart, the bloodstream in muscle is changed with the same period as the heart beat. Even in the case of no changing in the concentration of hemoglobin, there happens variation of blood stream (variation of blood volume). the variation of path length of the scattering light results in attenuation change in the measured scattering images. This change is related with the attenuation with a volume change of blood (so called AC component) and with the attenuation without a volume change of blood (so called DC component). Generally, the quantity of AC component in the attenuation by scattering is about 2% of the total attenuation. This component ($\Delta A$) is expressed by derivative of the total attenuation (A) with time in the region where an average path length (L) of arterial blood depends on only time.[8]

$$\Delta A = \frac{dA}{dt} \Delta t = \frac{dL}{dt} \cdot \sum_i \varepsilon_i \cdot c_i \cdot \Delta t \qquad (8)$$

where $\varepsilon_i, c_i$ are molar absorption coefficient and concentration of i-th component (i.e. $HHb$ or $O_2Hb$) respectively.

The temporal change of attenuation with time can appear even in attenuation of scattering light as well as in attenuation of penetrating light.

In case of measuring the scattering images of muscle by interval much shorter than the heart beat period, we intended to analyze the dynamic variation in arterial blood stream from time-dependent change of attenuation of the scattering light.

At first, the temporal change curve of attenuation was obtained from the scattering image frames and the photoplethysmogram was drawn from it.

By using the above sensor, p primary images were measured from human palm muscle and transformed into grey intensive images, from which attenuation images were obtained and composed into the attenuation image data cube.

To get more detailed information on the local positons of the measured images, the attenuation image data cube was converted to n fragmentary image data cube (simply, sub-data cube) by dividing each image of the attenuation image data cube into n fragment images,. From each sub-data cube, average attenuation value was calculated to obtain its attenuation change curve. By repating this procedure, n attenuation change curves can be obtained and used for local changes of photoplethysmogram.

In order to obtain AC component from the attenuation variation curve, a dynamic variation curve of attenuation ($\tilde{A}(t)$) was defined as follows.

$$\tilde{A}(t) = A(t) - \frac{1}{T}\int_{t_1}^{t_2} A(t)\,dt \cong A(t) - \frac{1}{p}\sum_{i=1}^{p} A(i) \qquad (9)$$

where $A(t)$ is attenuation at time $t$, $T = t_2 - t_1$, $p$ is a number of the acquired images.

In the dynamic variation curve of attenuation, there are not only the signals concerned with blood stream variation according to time, but also some noises from the measuring instruments and other sources. To eliminate the noises, Fourier transformation was applied. We gained a frequency power spectrum from the attenuation variation data by Fourier transformation and obtained photoplethysmogram by inverse Fourier transformation of it in the frequency area for the heart rate.

By using the above image sensor, we acquired 50 image frames on human palm muscle with the interval of 0.08s. By dividing each attenuation image into 32 fragment images, we obtained 32 attenuation change curves and 32 dynamic change curves of attenuation (Fig. 7(a) and Fig. 7(b)).

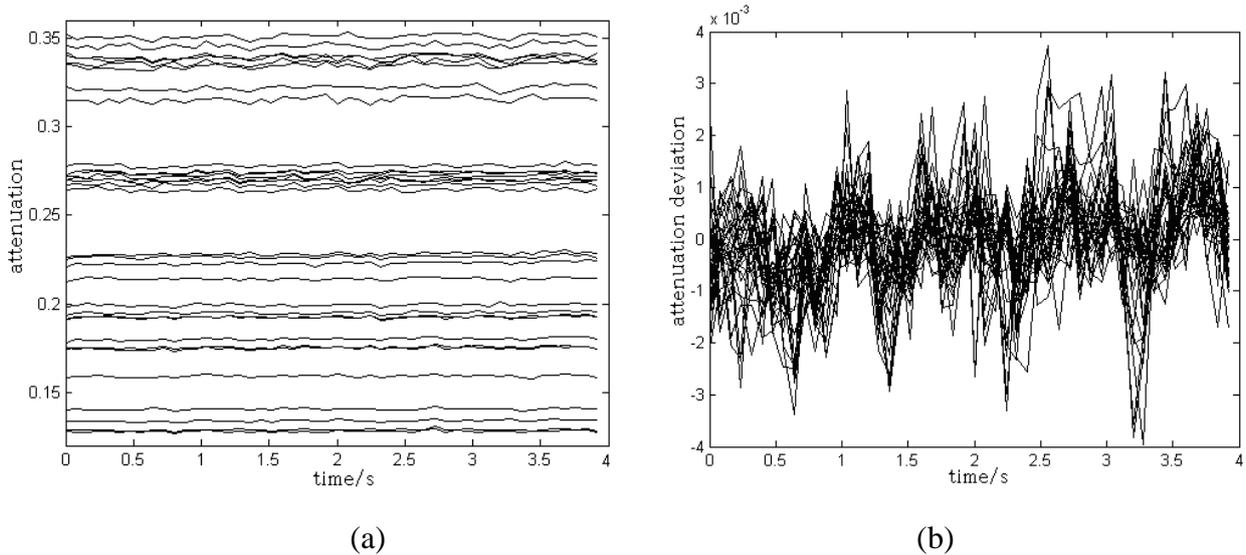

(a)    (b)

Fig. 7  Attenuation change curve (a) and dynamic change curve (b)

of attenuation obtained from 32-image data cube

Among those dynamic change curves, we selected the dynamic change curve with the most clear show of dynamic change property of the attenuation and processed it by Fourier transformation to obtain the frequency power spectrum (Fig 8), which was converted into the pure variation curve of blood stream by inverse Fourier transformation in the area of frequency 0.7~2s$^{-1}$ (Fig. 9).

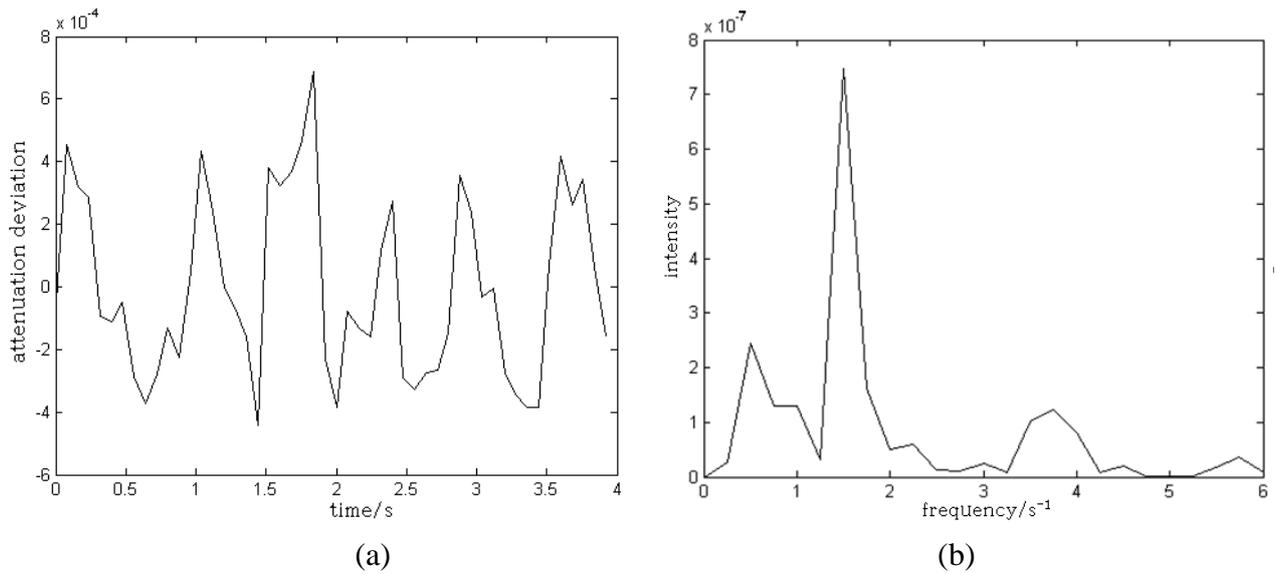

Fig. 8  The original signal (a) and frequency power spectrum (b) by Fourier transformation

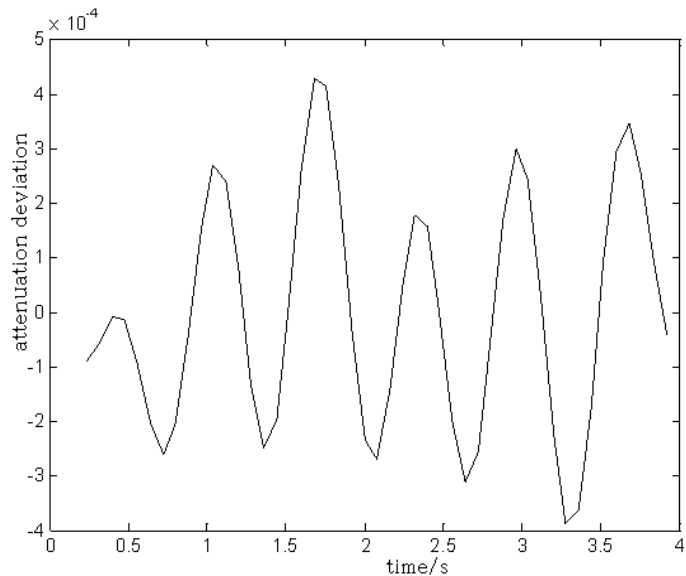

Fig. 9   The photoplethysmogram regenerated from the frequency power spectrum.

From this photoplethysmogram, blood stream through capillary vessel and heart rate can be interpreted. In Fig. 9 the attenuation deviation of the photoplethysmogram in the acquiring area was periodically changed according to the periodic change of blood stream. By using it, it is possible to determine the heat rate from the photoplethysmogram. For example, the heart rate from this photoplethysmogram was estimated to be 90.

Fig. 10 showed the photoplethysmograms and frequency spectra determined from the measured scattering images on palm muscles of several volunteerse 10.

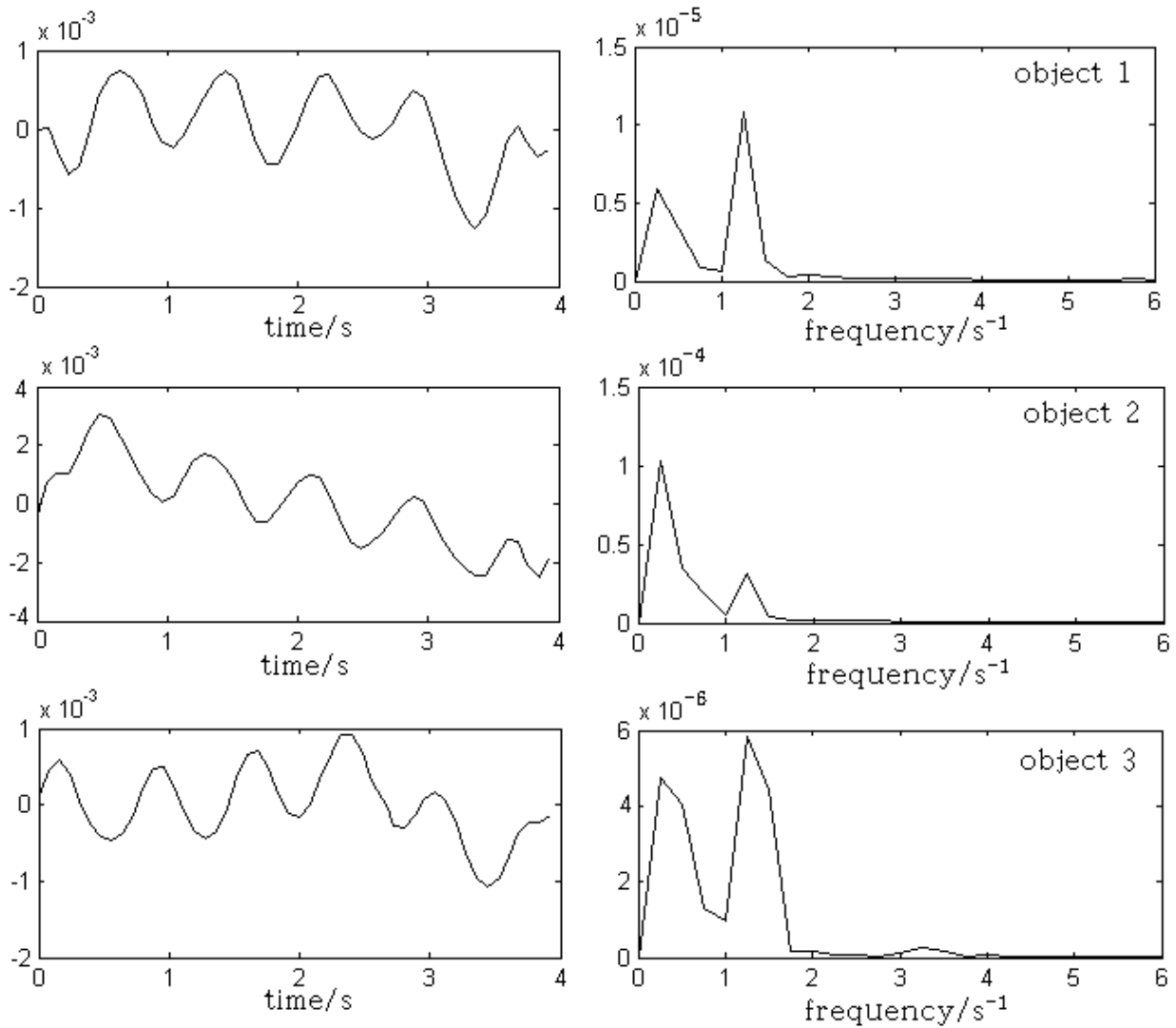

Fig. 10  The plethysmograms and frequency spectra from some volunteers

The shape of every maximum peak in the figure is consistent with that of the typical photoplethysmogram. In every plethysmogram the minimum point corresponds to the start point of the systole (or the end point of the diastole) and the maximum point corresponds to the start point of the diastole (or the end point of the systole).

**Conclusion**

The attenuation matrix and the separation matrix were determined from the NIR scattering images of human arm muscle and the extinction matrix was determined by using the modified Beer Lambert Law. From the extinction coefficient matrix, the distribution of oxygen saturation with separation were interpreted.

We determined the dynamic attenuation variation curve on the fragmentary image frames sensitive to blood stream from the temporal scattering image frames of palm muscle and then obtained the photoplethysmograms and heart rate by processing it with Fourier transformation and inverse Fourier transformation.